\documentclass[reprint, superscriptaddress, amsmath, amssymb, aps, prb, floatfix]{revtex4-2}

\usepackage{graphicx}
\usepackage{dcolumn}
\usepackage{bm}
\usepackage[T1]{fontenc}
\usepackage{lmodern}
\usepackage[utf8]{inputenc}
\usepackage{siunitx}
\usepackage[caption=false]{subfig}
\usepackage{xcolor}
\usepackage{hyperref}

\graphicspath{{.}{figs/}{XPS_normalized/}{D_appendix/}}
\begin{document}

\title{Effect of oxygen plasma treatment on the structural and superconducting properties of atomic-layer-deposited NbN thin films}

\author{Nitesh Yerra}
\email{nitesh.yerra@leibniz-ipht.de}
\affiliation{Leibniz Institute of Photonic Technology, 07745 Jena, Germany}
\affiliation{Institute for Solid State Physics, Friedrich-Schiller-Universit\"at Jena, 07743 Jena, Germany}

\author{Mustafa Teke}
\affiliation{Leibniz Institute of Photonic Technology, 07745 Jena, Germany}
\affiliation{Institute for Solid State Physics, Friedrich-Schiller-Universit\"at Jena, 07743 Jena, Germany}

\author{Felix Otto}
\affiliation{Institute for Solid State Physics, Friedrich-Schiller-Universit\"at Jena, 07743 Jena, Germany}

\author{Marco Diegel}
\affiliation{Leibniz Institute of Photonic Technology, 07745 Jena, Germany}

\author{Detlef Born}
\affiliation{Leibniz Institute of Photonic Technology, 07745 Jena, Germany}

\author{Uwe H\"ubner}
\affiliation{Leibniz Institute of Photonic Technology, 07745 Jena, Germany}

\author{Nan Du}
\affiliation{Leibniz Institute of Photonic Technology, 07745 Jena, Germany}
\affiliation{Institute for Solid State Physics, Friedrich-Schiller-Universit\"at Jena, 07743 Jena, Germany}

%
%

\author{Mario Ziegler}
\email{mario.luedicke-ziegler@leibniz-ipht.de}
\affiliation{Leibniz Institute of Photonic Technology, 07745 Jena, Germany}

\author{Sahitya Varma Vegesna}
\email{SahityaV.Vegesna@leibniz-ipht.de}
\affiliation{Leibniz Institute of Photonic Technology, 07745 Jena, Germany}
\affiliation{Institute for Solid State Physics, Friedrich-Schiller-Universit\"at Jena, 07743 Jena, Germany}

\author{Heidemarie Kr\"uger}
\email{heidemarie.krueger@leibniz-ipht.de}
\affiliation{Leibniz Institute of Photonic Technology, 07745 Jena, Germany}
\affiliation{Institute for Solid State Physics, Friedrich-Schiller-Universit\"at Jena, 07743 Jena, Germany}

\date{\today}

\begin{abstract}
Surface oxidation is a common concern for superconducting devices made of ultrathin NbN, yet which superconducting properties it affects has not been established for films grown by plasma-enhanced atomic layer deposition. We expose such NbN films of about 11.4\,nm thickness to an O$_2$ plasma for 0 to 60\,s, which oxidizes the surface without etching the film, and characterize the same sample series by X-ray diffraction, X-ray photoelectron spectroscopy (XPS), and transport measurements on 500\,nm wide nanowires. XPS of the surface and the underlying layer shows that the plasma converts the surface to Nb$_2$O$_5$, whereas the bulk of the treated film is chemically identical to that of the untreated film. The crystalline $\delta$-NbN remains intact. During the plasma exposure, the total film thickness rises and saturates between 40 and 60\,s, indicating a self-limiting growth of the surface oxide. The normal-state resistivity rises by 21\% (dc) and 25\% (optical) with the plasma dose, while the critical temperature and the transition width are unchanged. The switching current densities of 4.9--6.2\,MA/cm$^2$ at 4.2\,K show no systematic dependence on the plasma dose. Their temperature dependence follows the dirty-limit depairing current of Kupriyanov and Lukichev, and the extrapolated switching currents reach 39--46\% of the depairing current calculated from the optical resistivity and the diffusion coefficient, in agreement with the width dependence reported for nanowires from the same process. The oxygen signature is confined to temperatures near $T_\mathrm{c}$, where the supercurrent of the treated wires vanishes at least 0.5\,K earlier than that of the untreated wire, consistent with a partially oxidized surface layer that acts on the film by the proximity effect. At these doses, oxygen narrows the temperature range of the supercurrent from above but does not systematically reduce the low-temperature current-carrying capacity of the wires.
\end{abstract}

\maketitle

\section{Introduction}\label{sec:intro}

Ultrathin NbN films combine superconductivity at liquid-helium temperature with a short coherence length and a high normal-state resistivity. They remain superconducting down to thicknesses of a few nanometers, which suits them to nanoscale patterning and to large kinetic inductance. Films with thicknesses of a few nanometers form the active layer of superconducting nanowire single-photon detectors (SNSPDs)~\cite{Natarajan2012}, hot-electron bolometer mixers~\cite{Semenov2002}, and high-kinetic-inductance circuit elements~\cite{Niepce2019}. In addition, all-nitride qubit circuits employ NbN for junctions and wiring, motivated by its chemical stability against oxidation~\cite{Kim2021}. NbN interconnects with high kinetic inductance have also been integrated with aluminum Josephson junctions in a hybrid technology for quantum devices~\cite{Mutsenik2023}. Plasma-enhanced atomic layer deposition (PE-ALD) provides NbN films for such devices with monolayer thickness control and wafer-scale homogeneity~\cite{Ziegler_2013,Linzen2017,Knehr2019,knehr2021}. At these thicknesses the device performance follows the film properties directly.

Oxygen affects these films at every stage of their life. Air exposure forms a native oxide of 0.5 to 1\,nm on NbN within minutes~\cite{Semenov2009}, following the oxidation pathway to Nb$_2$O$_5$ above an intermediate niobium oxynitride~\cite{darlinski1987,jouve1996}. The sheet resistance of 5\,nm NbN films can rise by over 40\% after two years in atmosphere~\cite{Stepanov2024}. In sputtered NbN, oxygen can moreover penetrate deep into the film along the grain boundaries and degrade the superconductivity of the bulk~\cite{Henry2017}, and 3\,nm NbTiN films show no superconducting transition down to 3\,K unless the surface is passivated~\cite{Klein2026}. On the microwave side, the amorphous pentoxide is a dielectric that hosts two-level systems, which limit the quality factor of resonators and the coherence time of qubits~\cite{Bafia2024,Murthy2025,Gupta2026}, and surface oxidation could shift the operating frequency of kinetic-inductance circuits~\cite{Yang2024}. At the same time, O$_2$ plasma is deliberately applied to NbN during nanofabrication, for resist stripping and most recently as the oxidation half-cycle of atomic layer etching~\cite{Hossain2025}.

These observations/studies leave open where the oxygen acts to the film. Oxygen confined to a surface oxide reduces the superconducting cross section, whereas oxygen entering the grain boundaries or the lattice acts as bulk disorder and suppresses the critical temperature~\cite{Tyan1994,Cabanel1990,Henry2017}. Which case applies evidently depends on the microstructure of the film, and for PE-ALD NbN it has not been established. Natural aging is unsuited to answer this, since it is slow, poorly reproducible, and convolves oxidation with other degradation mechanisms. A controlled O$_2$ plasma dose separates the oxidation from these variables and accelerates it to practical timescales, so that the treated films serve as a proxy for aged films. If the oxide growth is self-limiting, the same treatment could moreover serve as a deliberate passivation of the surface, since the terminal oxide leaves nothing to oxidize further. The temperature-dependent switching current of a nanowire then distinguishes the two cases, since its low-temperature value probes the condensate while its behavior near $T_\mathrm{c}$ responds to perturbations on the scale of the diverging coherence length.

In this work, we establish a controlled oxygen incorporation in PE-ALD NbN films with a nominal thickness of 10\,nm (11.4\,nm by x-ray reflectivity, XRR) by O$_2$ plasma exposure, which grows a surface oxide without material removal, and determine its effect on the structural and superconducting transport properties. Films exposed for 0 to 60\,s are characterized by grazing-incidence x-ray diffraction (GIXRD) and x-ray photoelectron spectroscopy (XPS) before and after sputtering and by $R(T)$ and switching-current measurements on 500\,nm wide nanowires patterned from them. Those results will be compared with XRR and spectroscopic ellipsometry from colleagues on the same samples~\cite{Teke2026}. We find that the plasma oxidizes the film surface to Nb$_2$O$_5$ in a self-limiting way while the bulk of the film remains unchanged, so that it could be used as passivation. The critical temperature shows no degradation, and the switching current at 4.2\,K is nearly the same for all samples except B\_20s. The oxygen signature is confined to temperatures near $T_\mathrm{c}$, consistent with a weakened surface layer acting by the proximity effect~\cite{deGennes1964}. At these doses the oxygen therefore narrows the temperature range of the supercurrent but does not systematically reduce the low-temperature current-carrying capacity of the wires.

\section{Experimental Methods}\label{sec:methods}
\subsection{Film deposition and plasma treatment}\label{sec:deposition}

Four $10\times10$\,mm$^2$ chips were diced from a commercially available 100-mm boron-doped silicon wafer. On these substrates a \SI{4}{nm} Al$_2$O$_3$ buffer layer was deposited by atomic layer deposition (ALD) in an Oxford \emph{OpAL} reactor. This layer serves as an etch-stop during the subsequent patterning of the NbN. On top of this substrate, a \SI{10}{nm} NbN film was grown by PE-ALD~\cite{Ziegler_2013}. The reactor was operated at 5--300\,mTorr with a substrate table temperature of \SI{380}{\celsius}. The setup is installed inside a nitrogen-filled custom glovebox, which reduces oxygen contamination in the reactor during sample handling and processing. Each deposition cycle consisted of four steps. First, the substrate was exposed to the metal-organic precursor (tert-butylimido)tris(diethylamino)niobium (TBTDEN) for \SI{4}{s} to form an adsorbed monolayer. The chamber was then purged with argon to remove excess precursor. In the third step, a remote hydrogen plasma of \SI{300}{W} was applied for \SI{60}{s} to reduce and nitridize the adsorbed monolayer, followed by a final argon purge. The number of cycles was set to 188 to reach the target thickness of \SI{10}{nm}. The growth per cycle was 0.532\,\AA.

For the controlled incorporation of oxygen, three of the four samples were exposed to an O$_2$ plasma in a SENTECH Instruments GmbH SI~500-408 reactive ion etching (RIE) system with a capacitively coupled plasma source. The process parameters were \SI{4}{sccm} O$_2$ flow, \SI{6}{Pa} chamber pressure, and \SI{50}{W} rf power. Plasma exposure times of 20, 40, and \SI{60}{s} were applied, resulting in the samples B\_20s, C\_40s, and D\_60s, respectively. The as-deposited sample, which was not exposed to oxygen, is labeled A\_0s. For the XPS analysis, a duplicate pair of samples was prepared with the same number of ALD cycles and the same plasma process at the two exposures of 0 and 60\,s. This pair was grown on HF-cleaned Si substrates without the Al$_2$O$_3$ buffer layer to exclude substrate-related interference in the compositional analysis and was not patterned into wires. After the plasma treatment, the unpatterned films were characterized as described in Sec.~\ref{sec:characterization}. Subsequently, the samples were coated with AR-P~6200.04 electron-beam resist (Allresist GmbH) and patterned by electron-beam lithography (EBL) using a Vistec SB350-OS system. The structures were transferred into the NbN film by RIE with \SI{40}{sccm} CF$_4$ at \SI{100}{W} and \SI{9}{Pa}, carried out in the same RIE chamber used for the O$_2$ plasma treatment after a machine preconditioning step. The etch endpoint was monitored by \emph{in situ} reflectometry. The etch was stopped manually when the reflection signal changed from the NbN film to the substrate. The resulting etch times were 49, 80, 72, and 70\,s for A\_0s to D\_60s. The lithographic layout contained four-point measurement structures with a width of \SI{500}{nm} at a width-to-length ratio of 1:10.

\subsection{Characterization techniques}\label{sec:characterization}

\subsubsection{Structural and compositional analysis}
GIXRD measurements were performed on the unpatterned films using a PANalytical X'Pert PRO diffractometer with a Cu~K$\alpha$ source ($\lambda = 1.5406$\,\AA). The GIXRD patterns were recorded at a fixed incidence angle of $\omega = 2^{\circ}$ as a $2\theta$ scan from $10^{\circ}$ to $90^{\circ}$, repeated 45 times for statistics, to identify the crystalline phases and to evaluate the texture. Crystallite sizes were determined from a line-profile analysis of the (111) reflection using a Williamson-Hall approach in size-only mode. XPS was carried out with a PHOIBOS~150 hemispherical analyzer and monochromated Al~K$\alpha$ radiation ($h\nu = 1486.7$\,eV) from a FOCUS~500 monochromator with an XR50M X-ray source (SPECS Surface Nano Analysis GmbH). Survey and high-resolution core-level spectra (Nb~3d, N~1s, O~1s, C~1s) were recorded in the as-received state and after 15\,min of Ar$^+$ sputtering (700\,V, approximately \SI{4}{\micro\ampere\per\centi\metre\squared} at $5 \times 10^{-5}$\,mbar) to distinguish surface contamination from the composition at the center of the film. The film thicknesses were determined by XRR and spectroscopic ellipsometry was performed on the same unpatterned films to extract the Drude transport parameters (carrier density, mobility, and scattering time). The XRR analysis, the optical modeling and fitting procedure are described in detail in Ref.~\cite{Teke2026}.

\subsubsection{Electrical transport and critical current}
Resistance-temperature $R(T)$ curves were recorded in a four-point geometry on the patterned 500-nm-wide nanowire structures using a Keithley~2450 source-measure unit with a bias current of \SI{300}{\nano\ampere}. The critical temperature $T_\mathrm{c}$ was defined as the midpoint of the resistive transition (50\% of $R_\mathrm{max}$), and the transition width $\Delta T_\mathrm{c}$ as the temperature interval between the 90\% and 10\% resistance points. The measurement setup determines $T_\mathrm{c}$ with an accuracy of approximately 0.01\,K. Temperature-dependent critical current measurements were performed on the same 500-nm-wide, 5-\si{\micro\metre}-long nanowire bridges from \SI{4}{K} to \SI{13}{K} in \SI{0.25}{K} steps. The sample was mounted on a custom dipstick immersed in liquid helium. A resistive heater was installed on the underside of the sample breadboard to allow controlled warming above the helium bath temperature of \SI{4.2}{K}. The heater was mounted such that it is additionally coupled to the sample through the back side by a screw, which tightly connects both in a thermally conducting way. The temperature was set and stabilized with a Lakeshore~336 temperature controller before each measurement point. At each stabilized temperature, the critical current was determined from current-voltage ($I$-$V$) characteristics acquired with the Keithley~2450. The bias current was swept from 0 to \SI{400}{\micro\ampere} and back to 0 in \SI{10}{nA} steps. The switching current $I_\mathrm{sw}$ was defined as the bias current at which the nanowire undergoes a sudden transition from the superconducting zero-resistance state to the normal-conducting (resistive) state on the forward sweep. The retrapping current $I_\mathrm{r}$ was defined as the current at which the nanowire returns from the resistive state back to the superconducting state on the reverse sweep. Two consecutive sweeps were recorded at each temperature to assess the reproducibility of the measurement.

\section{Results}

\begin{figure}[!htb]
	\centering
	\includegraphics[width=\columnwidth]{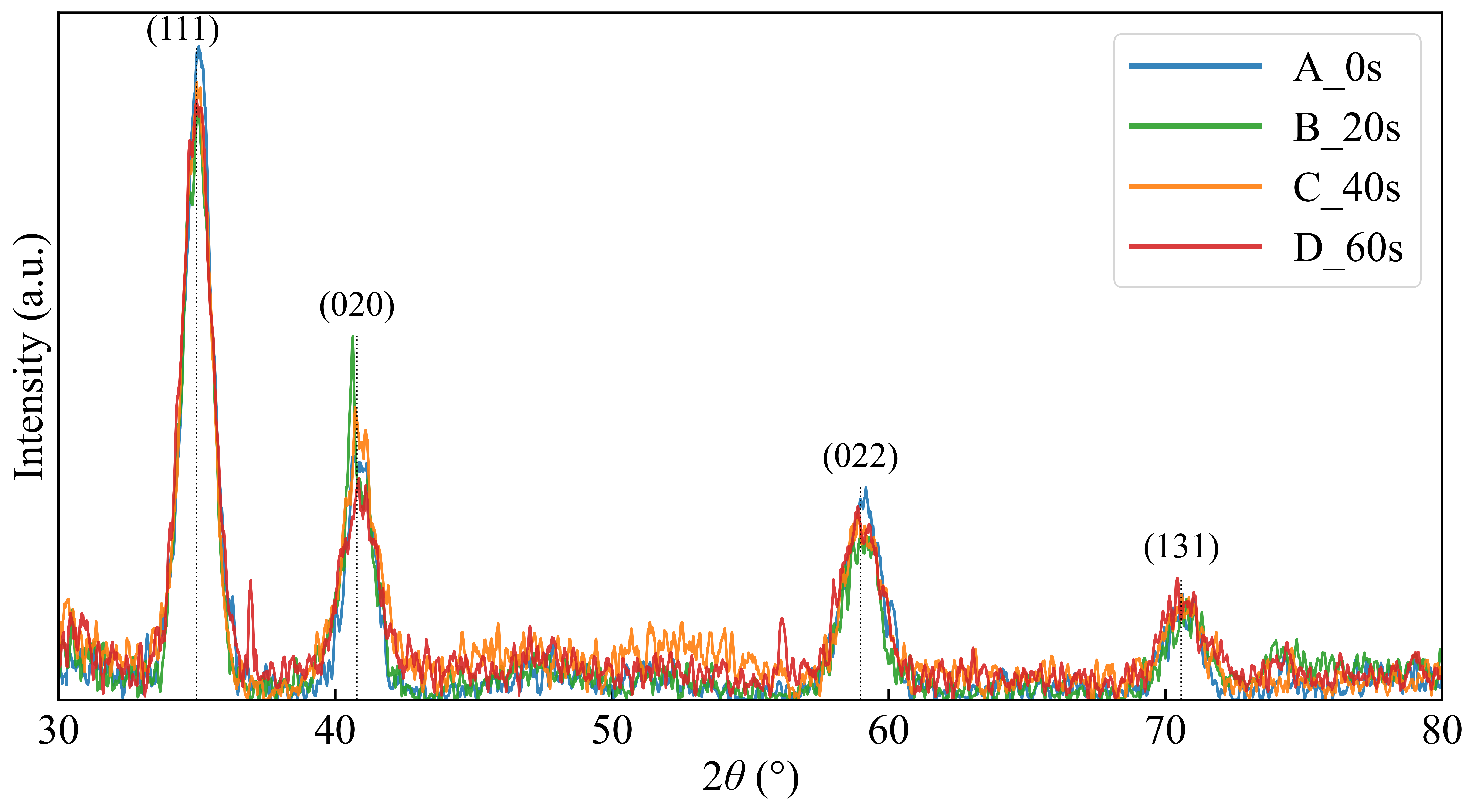}
	\caption{Background-subtracted GIXRD patterns of $\delta$-NbN thin films exposed to O$_2$ plasma for 0, 20, 40, and 60\,s (samples A\_0s, B\_20s, C\_40s, and D\_60s). The diffraction peaks correspond to the (111), (020), (022), and (131) reflections of cubic NbN. The raw GIXRD intensities were background-subtracted using a baseline fitted manually in the Malvern Panalytical HighScore Plus 4.9 powder diffraction software, smoothed with a Savitzky-Golay filter, and shown on a common arbitrary intensity scale (arb. units). Only the range of $30^{\circ}$--$80^{\circ}$ is shown as a zoomed view, since the region below is dominated by the substrate signal.\label{fig:xrdxrr}}
\end{figure}

\subsection{Material characterization}

\subsubsection{GIXRD analysis}

To evaluate the structural response to the plasma treatment, the four samples exposed for 0, 20, 40, and 60\,s were characterized by GIXRD. The GIXRD patterns of all four samples [Fig.~\ref{fig:xrdxrr}] confirm the formation of polycrystalline $\delta$-NbN (cubic)~\cite{Linzen2017,knehr2021}, with diffraction peaks corresponding to the (111), (020), (022), and (131) planes. The (111) reflection is the most intense one, indicating a preferred orientation along the close-packed direction of the cubic lattice. No additional diffraction peaks or peak splitting are observed across the three O$_2$ plasma exposure conditions. This indicates that the $\delta$-NbN phase is preserved and that no secondary phase crystallizes upon the plasma treatment. The crystallite size from the Williamson-Hall analysis remains approximately 6.5, 7.0, and 6.5\,nm for A\_0s, B\_20s, and C\_40s, respectively and decreases to 5.5\,nm for D\_60s. The microstrain term was set to zero because the microstrain uncertainties were found to be of the same order as the microstrain values themselves, rendering the combined size-plus-strain decomposition unreliable for these ultrathin films. The film thickness measured by XRR is 11.4, 11.8, 12.8, and 12.9\,nm from A\_0s to D\_60s~\cite{Teke2026}. It is nearly the same for the two longest exposures. This suggests that the growth of the surface layer saturates between 40 and 60\,s.

\begin{figure*}[t!]
\centering
\includegraphics[width=0.50\textwidth]{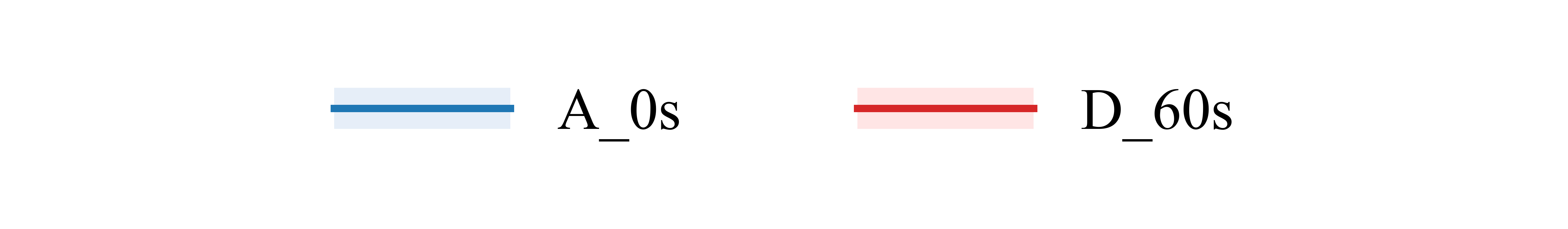}\\[8pt]
\textbf{Surface analysis}\\[4pt]
\begin{minipage}{0.24\textwidth}\centering\includegraphics[width=\textwidth]{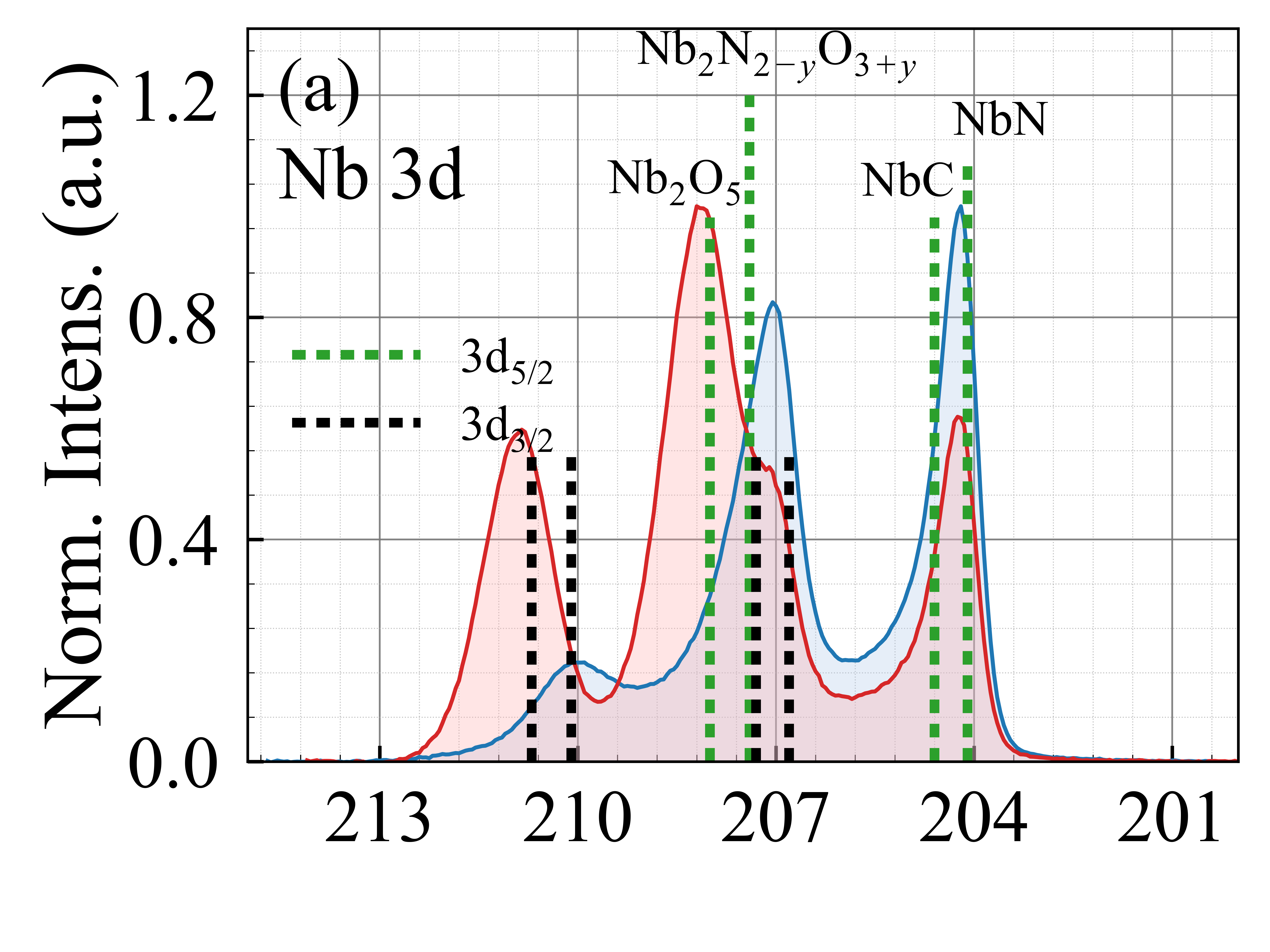}\end{minipage}%
\begin{minipage}{0.24\textwidth}\centering\includegraphics[width=\textwidth]{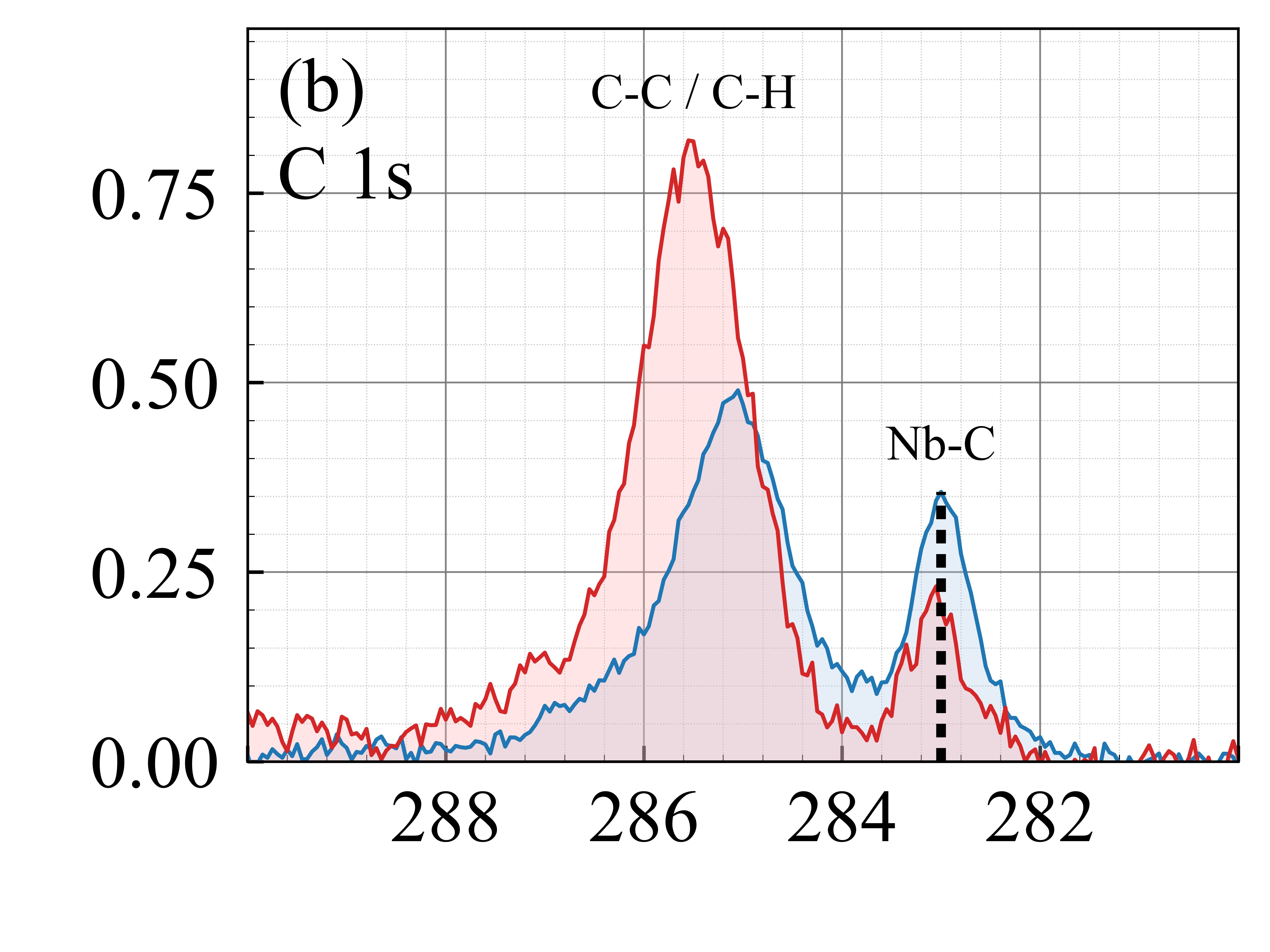}\end{minipage}%
\begin{minipage}{0.24\textwidth}\centering\includegraphics[width=\textwidth]{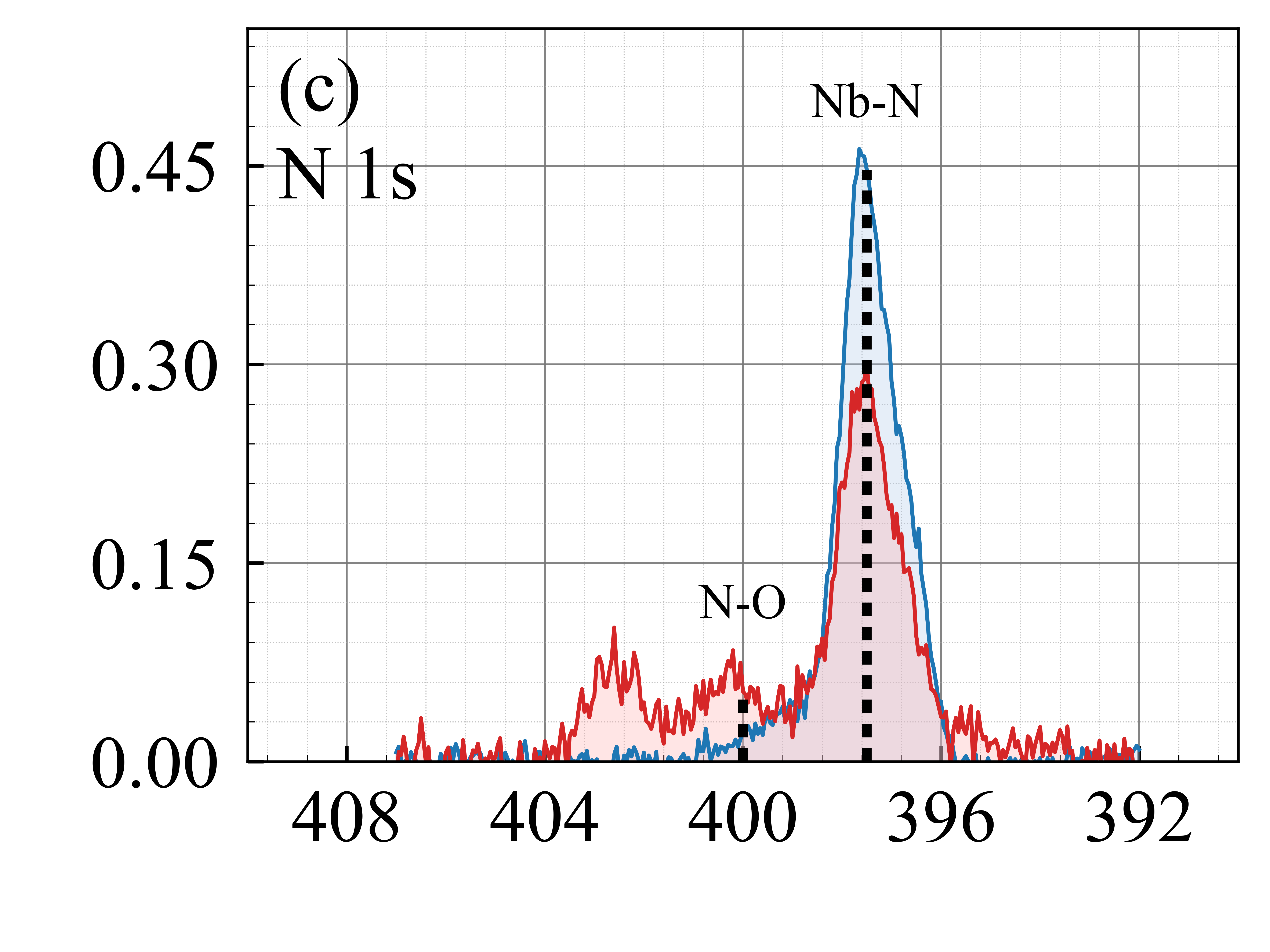}\end{minipage}%
\begin{minipage}{0.24\textwidth}\centering\includegraphics[width=\textwidth]{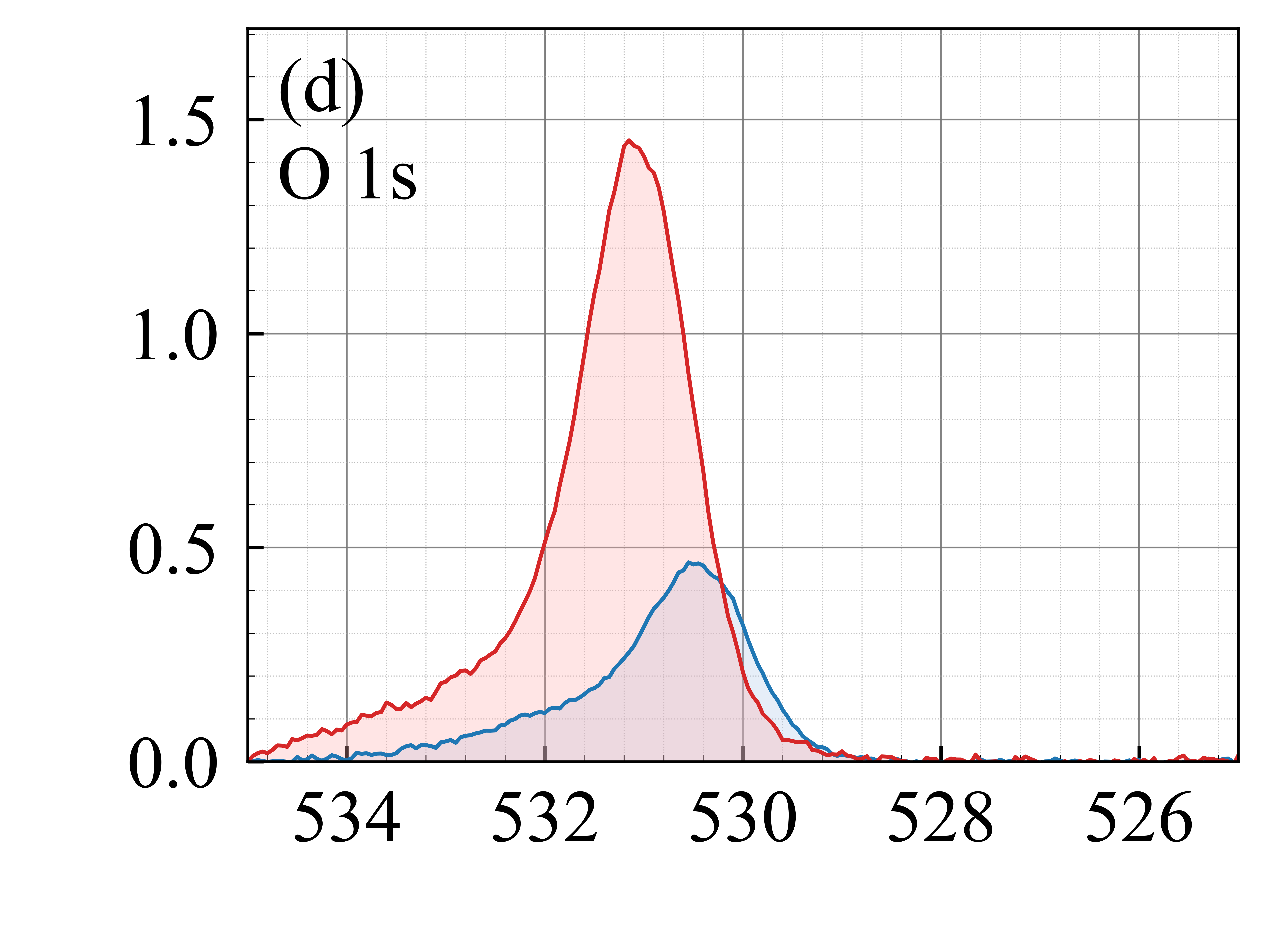}\end{minipage}\\[6pt]
\textbf{Center-of-film analysis}\\[4pt]
\begin{minipage}{0.24\textwidth}\centering\includegraphics[width=\textwidth]{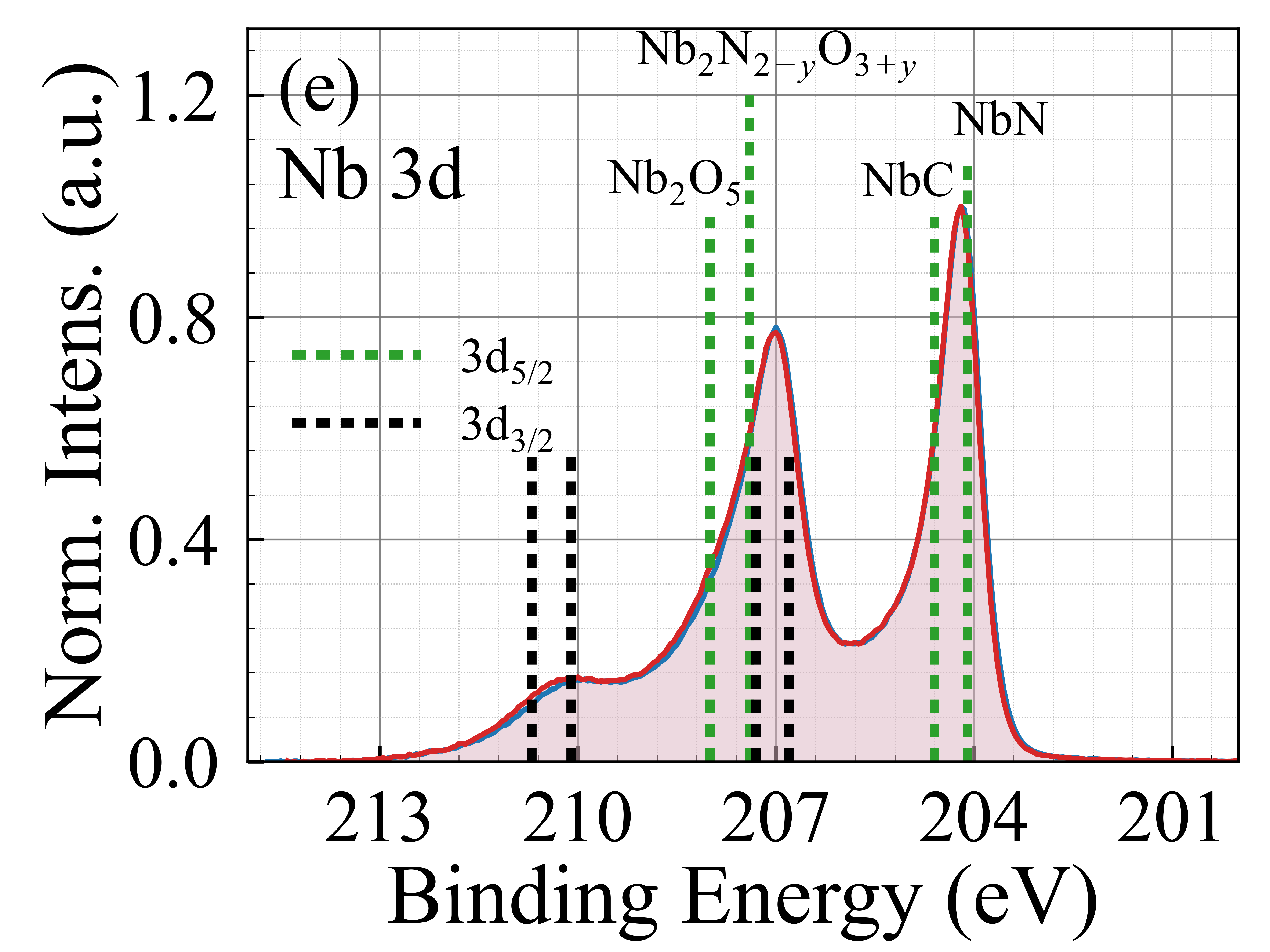}\end{minipage}%
\begin{minipage}{0.24\textwidth}\centering\includegraphics[width=\textwidth]{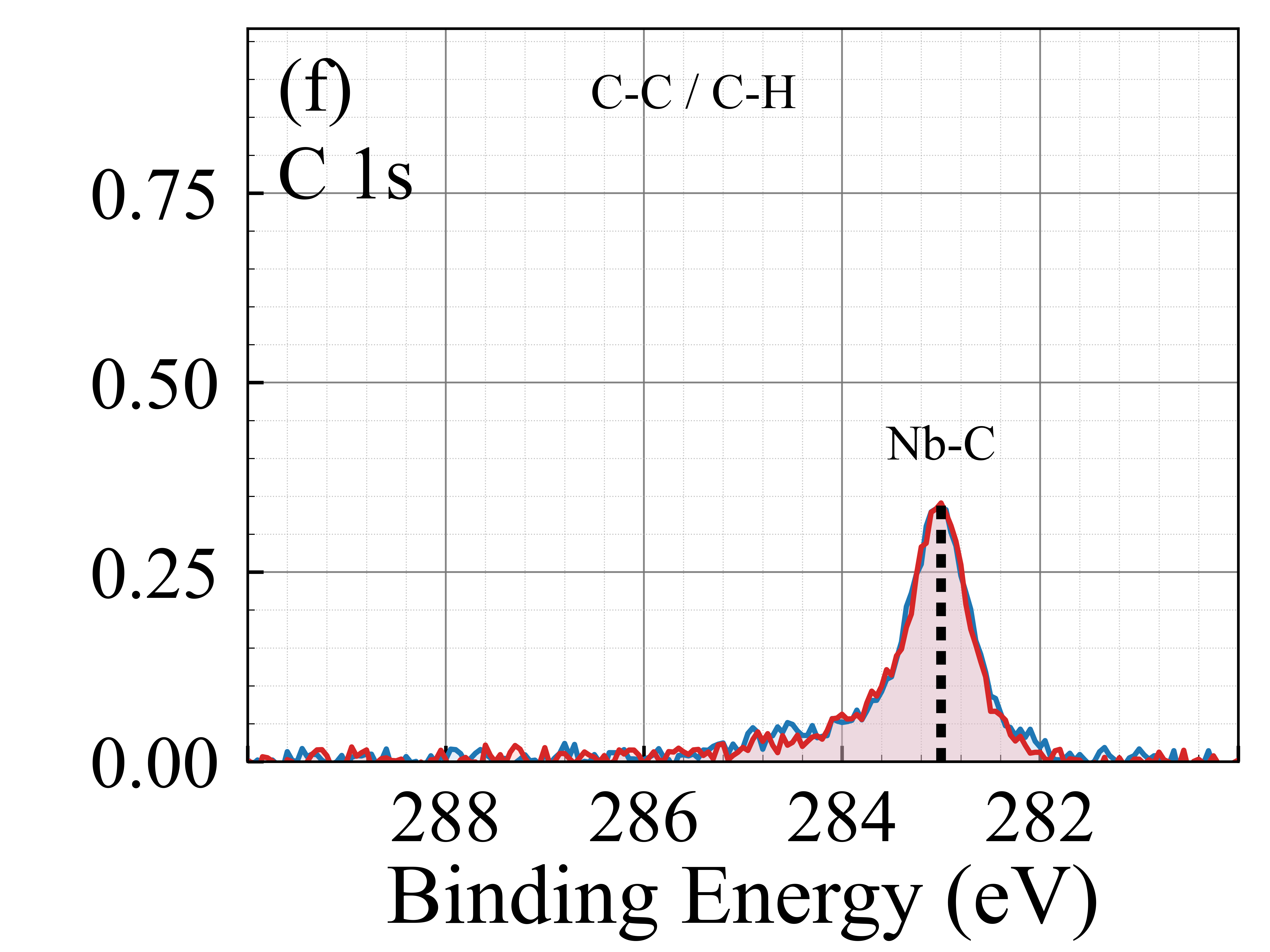}\end{minipage}%
\begin{minipage}{0.24\textwidth}\centering\includegraphics[width=\textwidth]{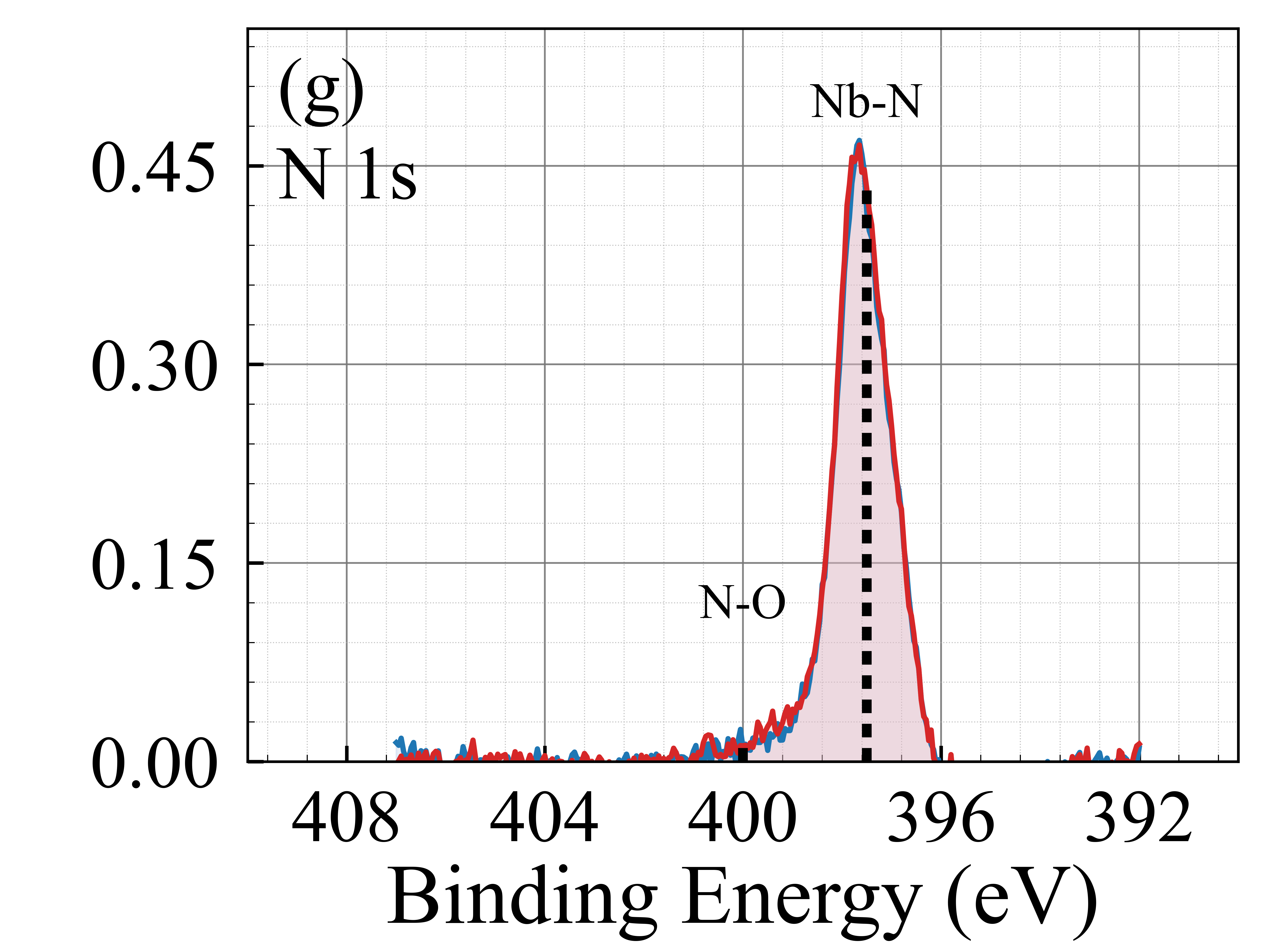}\end{minipage}%
\begin{minipage}{0.24\textwidth}\centering\includegraphics[width=\textwidth]{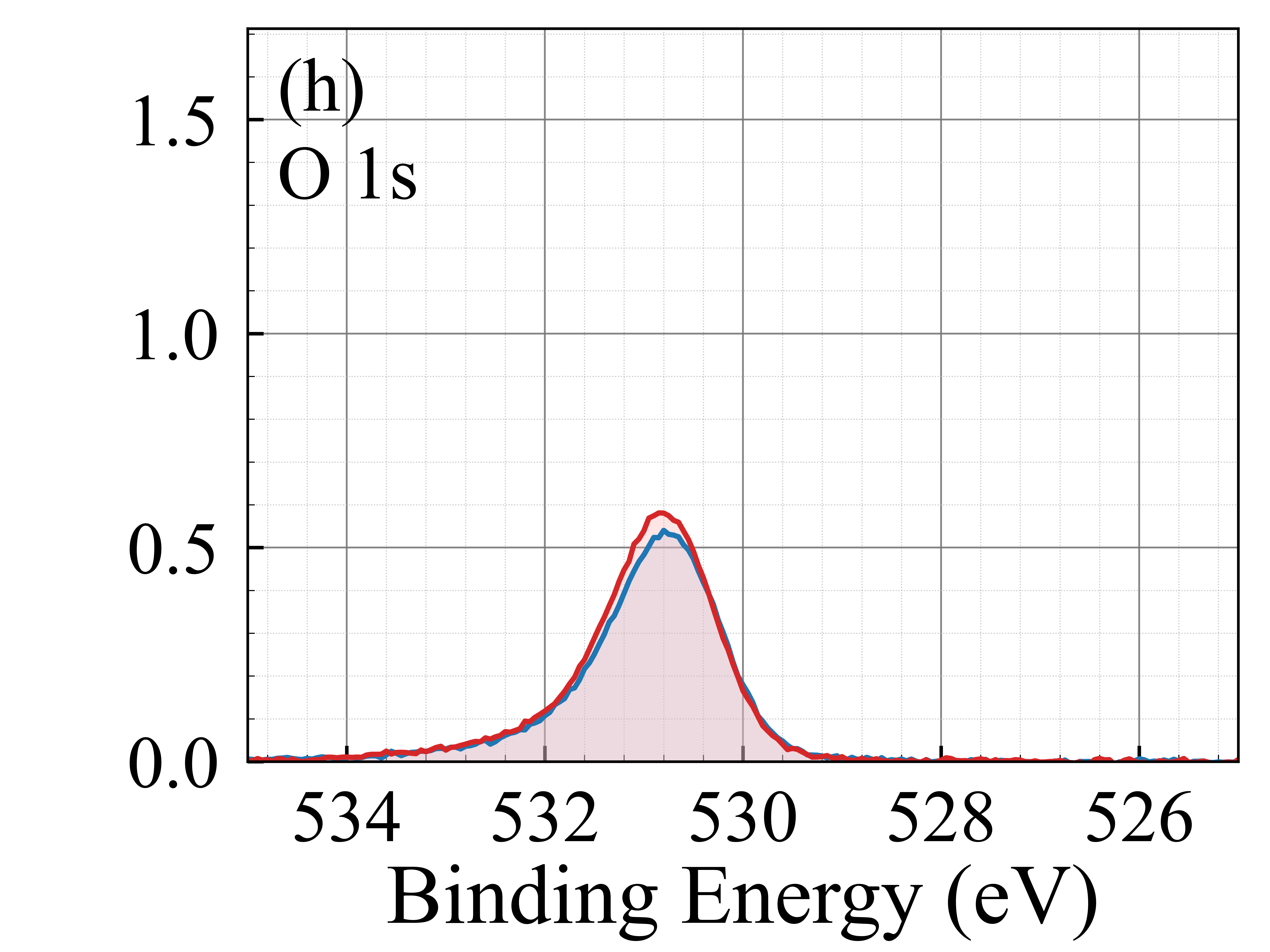}\end{minipage}
\caption{Background-subtracted XPS core-level spectra of the samples before and after sputtering, with gridlines added for direct intensity comparison. Top row, surface analysis (as-received): (a)~Nb~3d, (b)~C~1s, (c)~N~1s, (d)~O~1s. Bottom row, center-of-film analysis (after 15\,min Ar$^+$ sputtering): (e)~Nb~3d, (f)~C~1s, (g)~N~1s, (h)~O~1s. Blue: A\_0s; red: D\_60s. Dotted vertical lines mark the reference binding energies of the assigned components. In the Nb~3d panels the green lines are the $3d_{5/2}$ lines of NbN (204.1\,eV), NbC (204.6\,eV), Nb$_2$N$_{2-y}$O$_{3+y}$ (207.4\,eV), and Nb$_2$O$_5$ (208.0\,eV), and the black lines are the corresponding $3d_{3/2}$ partners, displaced by the spin-orbit splitting of 2.7\,eV and drawn at half height. In the C~1s and N~1s panels the lines mark Nb--C (283.0\,eV), Nb--N (397.5\,eV), and N--O (400.0\,eV). The spectra are normalized with the elemental sensitivity factors of~\cite{moulder1992} and scaled such that the maximum of the Nb~3d signal equals unity. For each core level, the surface and center-of-film panels therefore share the same intensity scale and can be compared directly.\label{fig:xps}}
\end{figure*}

\subsubsection{XPS analysis}

XPS was performed on the A\_0s and D\_60s samples to determine the chemical composition at the two limiting O$_2$ plasma exposures. The spectra were recorded on the duplicate, unpatterned sample pair described in Sec.~\ref{sec:deposition}. Each sample was measured in the as-received state, which reflects the film surface (\emph{surface analysis}), and after 15\,min of Ar$^+$ sputtering, which removes approximately 5\,nm and exposes the bulk (\emph{center-of-film analysis}). The background-subtracted core-level spectra are presented in Fig.~\ref{fig:xps}, with the surface analysis in the top row [Fig.~\ref{fig:xps}(a)--(d)] and the center-of-film analysis in the bottom row [Fig.~\ref{fig:xps}(e)--(h)].

Ar$^+$ bombardment of nitrides and oxides can introduce artifacts, including preferential sputtering of the lighter elements and ion-induced chemical reduction of surface oxides~\cite{jouve1996,Lubenchenko2018}. Thus, the sputtered spectra reflect the composition at the center of the film only approximately. Since both samples were sputtered under identical conditions, relative comparisons between A\_0s and D\_60s remain valid.

\paragraph{Nb 3d core-level spectra}
The Nb~3d region contains contributions from four niobium bonding environments, identified by their relative positions~\cite{darlinski1987} with the Nb--N $3d_{5/2}$ line observed at 204.1\,eV as reference and a spin-orbit splitting of 2.7\,eV~\cite{Ziegler2017}: Nb--N ($3d_{3/2}$ at 206.8\,eV), Nb--C at 204.6\,eV (207.3\,eV), the niobium oxynitride Nb$_2$N$_{2-y}$O$_{3+y}$ at 207.4\,eV (210.1\,eV), and Nb$_2$O$_5$ at 208.0\,eV (210.7\,eV). In the as-received state [Fig.~\ref{fig:xps}(a)], the Nb--N doublet is present in both samples and is more pronounced for A\_0s. The Nb--C contribution lies within the Nb--N peak. The peak of A\_0s at 207\,eV is a superposition of the $3d_{5/2}$ line of the niobium oxynitride with the $3d_{3/2}$ lines of Nb--N and Nb--C. For D\_60s, the dominant peak at 208.15\,eV corresponds to the $3d_{5/2}$ line of Nb$_2$O$_5$, a shoulder near 207\,eV to the niobium oxynitride superposed with the $3d_{3/2}$ lines of Nb--N and Nb--C, and the peak at 210.9\,eV mainly to the $3d_{3/2}$ line of Nb$_2$O$_5$ with a minor niobium oxynitride contribution. This indicates that the surface of A\_0s consists of NbN with some NbC and a niobium oxynitride layer without a resolved Nb$_2$O$_5$ component, whereas the surface of D\_60s is dominated by Nb$_2$O$_5$ with reduced NbN and minor NbC and niobium oxynitride contributions.

After sputtering [Fig.~\ref{fig:xps}(e)], both samples show the Nb--N doublet at 204.1\,eV as the dominant contribution, accompanied by Nb--C, and a peak at 207\,eV formed by the niobium oxynitride together with the $3d_{3/2}$ lines of Nb--N and Nb--C. No Nb$_2$O$_5$ component is resolved in the interior of either sample. The spectra before and after sputtering thus suggest that the carbide and the niobium oxynitride are constituents of the film interior, whereas Nb$_2$O$_5$ is confined to the surface.

\paragraph{C 1s core-level spectra}
The C~1s region [Fig.~\ref{fig:xps}(b) and (f)], in the as-received state of A\_0s [Fig.~\ref{fig:xps}(b)], a broad peak at 284--286\,eV is attributed to adventitious carbon (C--C, C--H, and C--O) that forms on all air-exposed surfaces~\cite{greczynski2017,moulder1992}. A distinct peak at approximately 283\,eV is assigned to niobium carbide (Nb--C)~\cite{moulder1992}. This feature persists after sputtering, indicating that it originates from the film. At the surface, the Nb--C peak of D\_60s is about 1.5 times weaker than that of A\_0s, whereas the peaks coincide after sputtering. The Nb--N signals of D\_60s are weaker than those of A\_0s by the same factor in the Nb~3d and N~1s regions [Fig.~\ref{fig:xps}(a) and (c)]. The suppression at the surface is therefore common to all core levels and is not specific to the carbide. The Nb--C peak thus points to niobium carbide in the film, a phase not resolved in the GIXRD patterns.

\paragraph{N 1s and O 1s core-level spectra}
The N~1s region [Fig.~\ref{fig:xps}(c) and (g)] is dominated by a peak at approximately 397.5\,eV, assigned to nitrogen in a Nb--N bonding environment~\cite{jouve1996}. In the as-received state, the Nb--N peak of D\_60s is weaker than that of A\_0s as part of the common surface suppression noted above. In addition, D\_60s shows a feature near 400\,eV corresponding to N--O bonding~\cite{Lubenchenko2018}. This indicates that oxygen partially substitutes nitrogen at the surface of D\_60s. After sputtering [Fig.~\ref{fig:xps}(g)], the Nb--N peaks of both samples coincide in position, width, and intensity, and the N--O feature has vanished. This indicates the same nitrogen content within the two films.

In the as-received state [Fig.~\ref{fig:xps}(d)], the O~1s signal of D\_60s is about 3 times more intense than that of A\_0s, and its maximum is shifted by about 0.6\,eV to higher binding energy, which correlates to Nb$_2$O$_5$ as identified in the Nb~3d spectra. Hydroxyl groups and adventitious C--O species adsorbed during air exposure contribute to the same envelope and are not resolved here~\cite{Idriss2021}. After sputtering [Fig.~\ref{fig:xps}(h)], the two profiles overlap in terms of position and intensity. This indicates the same oxygen content within the two films. This oxygen is bound to niobium and present at low concentration, as shown by the niobium oxynitride contribution that persists in the Nb~3d spectra.

\subsection{Electrical characterization}

\subsubsection{Above \texorpdfstring{$T_\mathrm{c}$}{Tc}}\label{sec:aboveTc}

\begin{table}[htb!]
\caption{Electrical transport parameters of the NbN films as a function of the O$_2$ plasma duration. The optical resistivity $\rho_\mathrm{opt}$ is obtained from the Drude fit of the ellipsometric spectra of the unpatterned films~\cite{Teke2026}.\label{tab:transport}}
\begin{ruledtabular}
\begin{tabular}{l c c c c c}
Sample & $T_\mathrm{c}$ & $\Delta T_\mathrm{c}$ & RRR & $\rho_\mathrm{opt}$ & $\rho_\mathrm{dc}$ \\
       & (K) & (K) & & (\si{\micro\ohm\centi\metre}) & (\si{\micro\ohm\centi\metre}) \\
\hline
A\_0s  & 12.13 & 0.93 & 0.904 & 213.2 & 537 \\
B\_20s & 11.77 & 0.93 & 0.921 & 239.8 & 592 \\
C\_40s & 12.19 & 0.98 & 0.918 & 251.9 & 639 \\
D\_60s & 11.77 & 0.94 & 0.916 & 266.9 & 648 \\
\end{tabular}
\end{ruledtabular}
\end{table}

Table~\ref{tab:transport} summarizes the electrical transport parameters extracted from the $R(T)$ measurements on the 500\,nm-wide nanowire structures together with the optical resistivity from spectroscopic ellipsometry~\cite{Teke2026}. The critical temperature lies within a narrow band of 11.7--12.2\,K for all four samples and shows no degradation with the plasma duration. The transition width remains essentially unchanged at $\Delta T_\mathrm{c} \approx 0.94$\,K. All samples exhibit an increase in resistance between 300\,K and $R_\mathrm{max}$. The residual resistivity ratio (RRR) is defined as $\mathrm{RRR}\equiv\rho(300\,\mathrm{K})/\rho_\mathrm{max}$, where $\rho_\mathrm{max}$ is the maximum resistivity reached just above the $T_\mathrm{c}$ onset~\cite{Pellan1990,Tyan1994}. This quantity is below unity for films with negative temperature coefficient of resistivity (TCR). The measured values lie between 0.904 and 0.921 (Table~\ref{tab:transport}). The RRR of the treated samples lies slightly closer to unity than that of the untreated film.
The dc resistivity of the wires increases from 537 to \SI{648}{\micro\ohm\centi\metre} with the plasma exposure (20.7\% increase). The optical resistivity of the unpatterned films increases from 213.2 to \SI{266.9}{\micro\ohm\centi\metre} (25.2\%). This increase is carried by a monotonic decrease of the carrier concentration by 14.2\% and of the mobility by 7.0\%, consistent with the Drude relation $\rho_\mathrm{opt} = 1/(Ne\mu)$~\cite{Teke2026}. Both resistivities rise most strongly within the first 20\,s of exposure. Beyond 40\,s the dc resistivity saturates, whereas the optical resistivity continues to rise. The ratio of the dc to the optical resistivity remains approximately constant at $2.49 \pm 0.06$ across all four samples.
\subsubsection{Below \texorpdfstring{$T_\mathrm{c}$}{Tc}}\label{sec:belowTc}

\begin{figure*}[t!]
  \centering
  \begin{minipage}{0.48\textwidth}
    \centering
    \includegraphics[width=\textwidth]{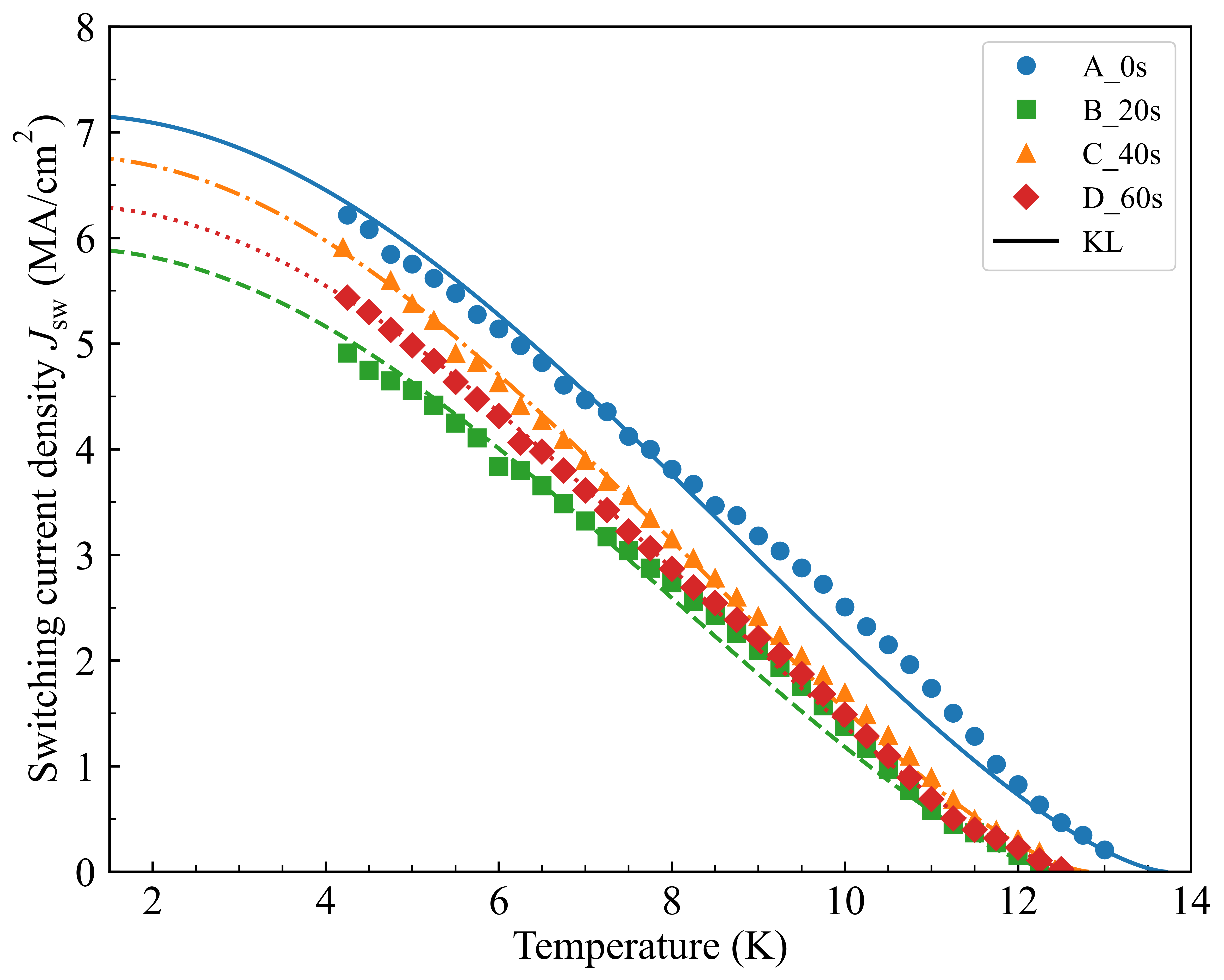}
  \end{minipage}\hfill
  \begin{minipage}{0.48\textwidth}
    \centering
    \includegraphics[width=\textwidth]{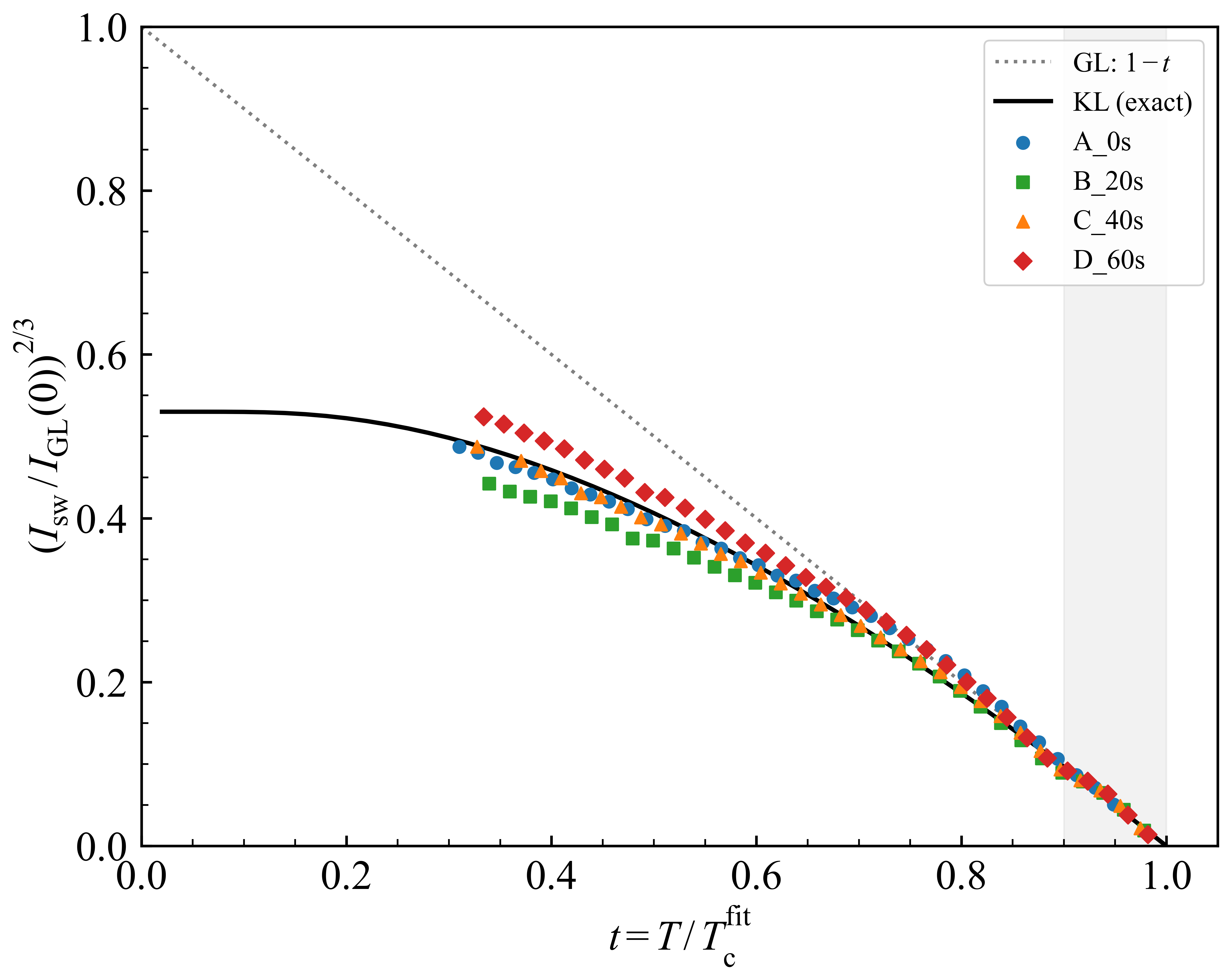}
  \end{minipage}
  \caption{(a)~Temperature dependence of the switching current density $J_\mathrm{sw}(T)$ for all four samples (symbols) together with dirty-limit Kupriyanov-Lukichev (KL) fits, Eqs.~(\ref{eq:kl}) and (\ref{eq:klcorr}) (lines). (b)~Reduced representation $(I_\mathrm{sw}/I_\mathrm{GL}(0))^{2/3}$ versus $t = T/T_\mathrm{c}^\mathrm{fit}$ following Ref.~\cite{Cirillo2007}. Dotted line: Ginzburg-Landau (GL) prediction $1-t$; solid line: exact dirty-limit KL solution of the Usadel equations, saturating at $0.386^{2/3} = 0.53$~\cite{ClemKogan2012}. A line fit over the finite window raises the expected saturation to 0.55--0.56. The shaded band marks the GL fit window $t \geq 0.9$ from which $I_\mathrm{GL}(0)$ and $T_\mathrm{c}^\mathrm{fit}$ are obtained.\label{fig:icjc}}
\end{figure*}

The oxygen plasma raises the normal-state resistivity systematically but leaves the resistive transition nearly unaffected. We now examine whether this robustness extends to the current-carrying capacity of the wires. The central measured quantity is the switching current $I_\mathrm{sw}$, the bias current at which the nanowire abruptly leaves the zero-resistance state on the forward branch of the hysteretic current-voltage characteristics. It is read directly from the recorded sweeps and involves no modeling. Two consecutive sweeps were recorded at each temperature. Premature switching can only lower the measured current, so the larger of the two values was retained. Points with $I_\mathrm{sw}$ below \SI{1}{\micro\ampere} were excluded, since such currents are comparable to the resolution of the sweep close to $T_\mathrm{c}$. For presentation the currents are converted to current densities $J_\mathrm{sw} = I_\mathrm{sw}/(wd)$ with the lithographic width $w = 500$\,nm and the XRR film thicknesses. The thickness of the plasma-modified surface layer cannot be determined from our data, since XRR yields only the total film thickness, which is therefore used for all current densities. At the base temperature of 4.2\,K the measured switching currents are 354, 290, 378, and \SI{351}{\micro\ampere}, corresponding to 6.2, 4.9, 5.9, and 5.4\,MA/cm$^2$. At this temperature the current density of A\_0s is the highest; C\_40s lies 5\%, D\_60s 13\%, and B\_20s 21\% lower, without a trend in dose. A nonsuperconducting surface layer would lower the current in proportion to its thickness. Its effect is therefore not resolved against this scatter and is treated as a common uncertainty in the following. Figure~\ref{fig:icjc}(a) shows $J_\mathrm{sw}(T)$ for all four samples. Above approximately 7\,K the three oxygen-exposed samples fall progressively below A\_0s, and their supercurrent vanishes between 12.3 and 12.5\,K, whereas A\_0s still carries a supercurrent at 13.0\,K, the upper limit of the measurement. The oxygen signature thus resides in the temperature dependence of the switching current rather than in its low-temperature magnitude.

The zero-temperature gap is obtained from the strong-coupling relation $\Delta_0 = \beta_0\,k_\mathrm{B}T_\mathrm{c}$ with $\beta_0 = 2.0$, measured by terahertz spectroscopy on films from the same deposition process~\cite{Bolle2026}. Applying it to each sample's own $T_\mathrm{c}$ keeps the gap per-sample and assumes that the coupling strength is unaffected by the plasma, consistent with the unchanged film interior. Comparable ratios were employed for nanowires ($\beta_0 = 2.05$)~\cite{Knehr2019} and measured by tunneling spectroscopy on 4--5\,nm films from the same process ($\beta_0 = 2.14$)~\cite{Lorenz2026}. The resistivity enters as the optical value $\rho_\mathrm{opt}$ of Table~\ref{tab:transport}, since it describes the conducting film itself, whereas $\rho_\mathrm{dc}$ of the wires additionally contains the grain-boundary barriers and the fabrication-related contributions discussed below. The electron diffusion coefficient, $D = 0.53$\,cm$^2$/s, was determined from the upper critical field of an untreated film deposited with the same process (Appendix~\ref{app:D}). The same value is used for all four samples, since the XPS spectra after sputtering show the film below the modified surface layer to be unchanged by the plasma. The coherence length is $\xi = \sqrt{\hbar D/\Delta_0} = 4.07$--$4.15$\,nm, and the penetration depth is $\lambda = \sqrt{\hbar\rho_\mathrm{opt}/(\pi\mu_0\Delta_0)} = 412$--$468$\,nm (A\_0s to D\_60s). The high resistivity and the small diffusion coefficient place these films in the dirty limit, in which both expressions hold.

For such superconductors the appropriate temperature dependence over the full temperature range is the dirty-limit depairing current of Kupriyanov and Lukichev (KL)~\cite{Kupriyanov1980}. We fit the measured switching currents with its closed-form approximation, Eqs.~(\ref{eq:kl}) and (\ref{eq:klcorr})~\cite{Charaev2017,KnehrDiss},
\begin{equation}
J_\mathrm{sw}(T) = J_\mathrm{sw}(0)\,(1-t)^{3/2}\,\frac{\mathrm{KL}(t)}{\mathrm{KL}(0)},
\qquad t = T/T_\mathrm{c}^\mathrm{fit},
\label{eq:kl}
\end{equation}
\begin{equation}
\mathrm{KL}(t) = 1.87 - \frac{1.46}{1+(0.7\,t)^{1.12}}.
\label{eq:klcorr}
\end{equation}
Here $\mathrm{KL}(t)$ approximates the exact dirty-limit solution, with $\mathrm{KL}(0) = 0.41$, and $J_\mathrm{sw}(0)$ is the zero-temperature amplitude of the fit. The temperature scale $T_\mathrm{c}^\mathrm{fit}$ is obtained from the data alone. For each sample a straight line is fitted to $I_\mathrm{sw}^{2/3}$ versus $T$ using only the near-transition points $T \geq 0.9\,T_\mathrm{c}^\mathrm{fit}$ (4--5 points per sample), and its intercept with the temperature axis defines $T_\mathrm{c}^\mathrm{fit}$. It is held fixed in the fit, so that $J_\mathrm{sw}(0)$ is the only free parameter. The fits describe the measured $J_\mathrm{sw}(T)$ with $R^2 = 0.987$--$0.999$ [lines in Fig.~\ref{fig:icjc}(a)] and yield the zero-temperature amplitudes listed in Table~\ref{tab:eta}. $J_\mathrm{sw}(0)$ is an extrapolated fit amplitude, not a measured value. $T_\mathrm{c}^\mathrm{fit}$ exceeds the resistive midpoint by 0.6--1.6\,K. A comparable offset between the supercurrent and resistive temperature scales has been reported for disordered NbTiN wires~\cite{Hortensius2012}.

The corresponding theoretical maximum is the depairing current. In the exact dirty limit it reads~\cite{ClemKogan2012}
\begin{equation}
J_\mathrm{dep}(0) = 0.74\,\frac{\Delta_0^{3/2}}{e\,\rho_\mathrm{opt}\sqrt{\hbar D}},
\label{eq:jdep}
\end{equation}
and is calculated entirely from the material parameters above. Its temperature dependence follows the same dirty-limit form, Eqs.~(\ref{eq:kl}) and (\ref{eq:klcorr}), with $t = T/T_\mathrm{c}$. The resulting depairing currents lie between 13.6 and 17.8\,MA/cm$^2$, i.e., between 875 and \SI{1012}{\micro\ampere} (Table~\ref{tab:eta}).
To test the temperature dependence of the switching current without material parameters we adopt the reduced representation shown in Fig.~\ref{fig:icjc}(b), introduced for superconductor/ferromagnet bilayers~\cite{Cirillo2007}. The straight line through the near-transition points that defines $T_\mathrm{c}^\mathrm{fit}$ is the near-$T_\mathrm{c}$ limit of Eq.~(\ref{eq:kl}), the Ginzburg-Landau (GL) law $I_\mathrm{sw}(T) = I_\mathrm{GL}(0)\,(1-t)^{3/2}$ with $I_\mathrm{GL}(0) = I_\mathrm{sw}(0)/\mathrm{KL}(0)$, to which all candidate theories reduce near $T_\mathrm{c}$. In the coordinates $(I_\mathrm{sw}/I_\mathrm{GL}(0))^{2/3}$ versus $t = T/T_\mathrm{c}^\mathrm{fit}$, this law is the straight line $1-t$, and the exact dirty-limit KL solution of the Usadel equations saturates at 0.53~\cite{Kupriyanov1980,ClemKogan2012}. A straight line fitted over a finite near-$T_\mathrm{c}$ window does not yet reach the asymptotic slope. A\_0s and C\_40s follow the dirty-limit prediction and agree with it within 4\%. B\_20s lies 9\% below it, and D\_60s lies 5\% above it. All four deviations remain within 1.2 standard deviations of the extrapolation uncertainty. A saturation above the prediction means that the near-$T_\mathrm{c}$ part of the curve is suppressed relative to its low-temperature part. The deviation is largest for D\_60s, the sample with the largest plasma dose.

The absolute scale of the switching currents is determined by the depairing efficiency $\eta = I_\mathrm{sw}(0)/I_\mathrm{dep}(0)$, the fitted zero-temperature amplitude divided by the calculated maximum. The efficiencies lie between 0.39 and 0.46 (Table~\ref{tab:eta}), so the wires reach less than half of their intrinsic limit. They show no systematic dependence on the oxygen dose. The collapse test normalizes to the near-$T_\mathrm{c}$ extrapolation of the data and requires no material parameters, whereas $\eta$ normalizes to the calculated depairing current. The two tests are therefore complementary.

\begin{table}[ht]
\caption{Fitted switching-current amplitudes, calculated depairing currents, and depairing efficiencies $\eta = I_\mathrm{sw}(0)/I_\mathrm{dep}(0)$ of the 500\,nm wires. All entries are zero-temperature values, the amplitudes extrapolated from the fits and the depairing currents calculated, evaluated with $\Delta_0 = 2.0\,k_\mathrm{B}T_\mathrm{c}$ and $D = 0.53$\,cm$^2$/s (see text).\label{tab:eta}}
\begin{ruledtabular}
\begin{tabular}{l c c c c c}
Sample & $I_\mathrm{sw}(0)$ & $J_\mathrm{sw}(0)$ & $I_\mathrm{dep}(0)$ & $J_\mathrm{dep}(0)$ & $\eta$ \\
       & (\si{\micro\ampere}) & (MA/cm$^2$) & (\si{\micro\ampere}) & (MA/cm$^2$) & \\
\hline
A\_0s  & 407.9 & 7.16 & 1012 & 17.8 & 0.40 \\
B\_20s & 347.6 & 5.89 & 891 & 15.1 & 0.39 \\
C\_40s & 432.6 & 6.76 & 969 & 15.1 & 0.45 \\
D\_60s & 405.9 & 6.29 & 875 & 13.6 & 0.46 \\
\end{tabular}
\end{ruledtabular}
\end{table}

\section{Discussion}\label{sec:discussion}

\subsection{Location of the plasma-induced oxygen}

The XPS spectra before and after sputtering locate the plasma-induced oxygen at the surface. The replacement of the oxynitride-terminated surface by Nb$_2$O$_5$ [Figs.~\ref{fig:xps}(a) and \ref{fig:xps}(d)] follows the established oxidation pathway of NbN. Nb$_2$O$_5$ forms the outermost oxide and carries the growth of thicker oxides, underlain by the dielectric niobium oxynitride Nb$_2$N$_{2-y}$O$_{3+y}$~\cite{darlinski1987}. For oxidation in air the oxynitride reaches a mean thickness of about 0.4\,nm and does not grow further, whereas plasma oxidation has been reported to produce a homogeneous Nb$_2$O$_5$ layer beyond 3\,nm~\cite{darlinski1987}. Our spectra before and after sputtering are consistent with a plasma-thickened Nb$_2$O$_5$ layer over an oxynitride that remains in place. Its thickness under plasma exposure is not established here. Nb$_2$O$_5$ is one of the thermodynamically stable phases of the Nb--O system~\cite{Nico2016} and was not detected at the A\_0s sample. This suggests that its formation requires a stronger oxidation than the exposure to air or to residual oxygen during the deposition. A comparative oxidation study of sputtered NbN films supports this. Under controlled oxidation at up to \SI{200}{\celsius} the oxide on NbN saturates near 2\,nm, air-stored films carry about 3\,nm, and only plasma oxidation grows the oxide beyond 8\,nm, with nitrogen leaving the oxidized layer~\cite{Ermolieff1985}. The plasma therefore drives the surface within one minute to a state at or beyond the endpoint of natural aging. Spectroscopic ellipsometry on the same films shows an additional absorption band near 4.9\,eV that appears only after the plasma treatment and has been assigned to the oxidized surface region~\cite{Teke2026}.

The surface oxide leaves no trace in GIXRD [Fig.~\ref{fig:xrdxrr}]. An x-ray amorphous surface layer does not contribute to the diffraction pattern, which explains why no additional reflections appear although the film thickness increases. An increasing oxygen content in Nb--O--N films has been reported to change the structure from a columnar crystalline microstructure toward x-ray amorphous niobium oxynitride phases~\cite{Fenker2006}. The smaller crystallite size of D\_60s may reflect a reduced coherent NbN thickness after the surface oxide has formed.

In granular NbN, the grain boundaries host dielectric Nb$_2$O$_5$ and niobium oxynitride~\cite{darlinski1987}. With crystallites of 5.5--7\,nm, our films provide a dense network of such boundaries. The niobium oxynitride found in the film is therefore a grain-boundary oxide formed during growth. Since its contribution to the sputtered Nb~3d spectra is weak [Fig.~\ref{fig:xps}(e)], the oxynitride decorates the grain boundaries rather than enclosing the grains.

The two impurities (O, C) have different origins. The C~1s spectra show niobium carbide, whose carbon is supplied by the ligands of the metal-organic precursor and remains in the film when the hydrogen plasma does not remove them completely~\cite{Ziegler2017,knehr2021}. Cubic NbC is isostructural with $\delta$-NbN and fully miscible with it, and the lattice parameters of the two end members are closely spaced ($a_{\mathrm{NbN}} = 4.39$\,\AA~\cite{Brauer1961}, $a_{\mathrm{NbC}} = 4.47$\,\AA~\cite{Storms1959}). The reflections of the two compounds therefore overlap, which explains why the carbide appears in XPS but not in GIXRD. The carbon most likely occupies the anion sublattice of the rock-salt lattice rather than forming a separate phase. The other impurity is oxygen. The precursor itself contains no oxygen. The oxygen seen in the O~1s spectra is therefore most likely incorporated from residual water and oxygen in the deposition chamber. For films from the same process, the Nb~3d intensity between the Nb--N and the Nb$_2$O$_5$ positions was previously assigned as a whole to a mixture of niobium oxide, oxynitride, and carboxide species~\cite{Ziegler2017}. The comparison before and after sputtering separates this mixture by location and is consistent with that picture. The plasma therefore modifies only the surface. It changes neither the grain boundaries nor the interior of the film.

\subsection{Normal-state transport}

The location of the oxygen at the surface constrains what the normal-state transport can show. Neither the NbN lattice nor the grain boundaries are affected, so the resistivity should rise while the critical temperature and the conduction mechanism remain unchanged. The measured transport follows this expectation (Table~\ref{tab:transport}). This robustness is consistent with the insensitivity of the critical temperature of a conventional $s$-wave superconductor to nonmagnetic impurity scattering~\cite{Anderson1959}. Magnetic pair breaking is likewise not expected. The unchanged transition width indicates that the plasma introduces no macroscopic secondary phase or nonsuperconducting region, in agreement with the single-phase $\delta$-NbN observed by GIXRD. Disorder does suppress $T_\mathrm{c}$ in granular NbN, where a resistivity increase by nearly an order of magnitude lowers it from 11.7 to 5.4\,K~\cite{Tyan1994}. Extrapolating that suppression to the 20.7\% increase observed here predicts a shift of about 0.06\,K, well below the sample-to-sample scatter.

The negative TCR is characteristic of strongly disordered NbN. In granular films it has been attributed to transport limited by the grain boundaries, either by intergrain tunneling~\cite{Nigro1988} or by a strongly reduced electron transmission across the boundaries~\cite{Tyan1994}. In NbN these barriers are the grain-boundary dielectrics identified above, which cause the RRR below unity~\cite{darlinski1987}. The dielectric itself carries no current. The conduction path is metallic within the NbN grains, and the boundary oxide acts as a thin barrier that the carriers must cross. The optical response probes the carriers inside the grains, whereas the dc measurement crosses the boundaries as well. The ratio $\rho_\mathrm{dc}/\rho_\mathrm{opt}$ is therefore a measure of the barrier contribution. That contribution is moderate. The resistivity rises by only 8 to 11\% between 300\,K and $T_\mathrm{c}$, and the resistive transition is a single step of 0.94\,K. If a dielectric shell enclosed the grains completely, the wire would turn fully superconducting only at a lower temperature governed by the Josephson coupling between the grains~\cite{Pellan1990}. The grains in our films are therefore well connected, and the boundary oxide adds a series resistance rather than isolating them.

The shift of the RRR toward unity for the treated samples is expected, since a constant added resistivity moves the ratio toward unity. The measured resistivity increase of 20.7\% predicts $\mathrm{RRR} \approx 0.92$, in agreement with the observed values. The plasma treatment adds elastic scattering but does not change the transport mechanism. The optical response shows the same behavior. The disorder-related infrared absorption of the films weakens with the dose but stays at the same energy, a sign of additional scattering at the modified surface rather than of a different conduction mechanism~\cite{Teke2026}.

The resistivity increase appears in the wires and in the optical response of the unpatterned films alike. The optical model treats the NbN layer as one homogeneous medium of the XRR thickness, so the Drude parameters are an average over the entire film including the oxidized surface~\cite{Teke2026}. The decrease of $N$ and $\mu$ with the plasma dose is explained there by oxygen bonded to niobium at the surface, which removes a few carriers and scatters the remaining ones~\cite{Teke2026}. The increase of both resistivities is therefore consistent with the surface modification, while the film below remains unchanged. The dc resistivity saturates together with the film thickness, consistent with the surface oxide as the source of the added scattering. The grain boundaries are not modified by the plasma, as established above. The surface oxide therefore adds scattering without altering the intergrain barriers, in contrast to granular sputtered NbN in which incorporated oxygen progressively strengthens the insulating barriers between the grains~\cite{Cabanel1990}.

The ratio of the dc to the optical resistivity, $2.49 \pm 0.06$, quantifies the contributions absent from the optical value, namely the grain-boundary barriers, fabrication-related contributions such as etch damage at the wire edges and deviations of the cross section from the lithographic layout, and the reduced dc conductivity of carriers whose spectral weight is shifted to finite frequency by disorder~\cite{Teke2026}. Since the chips were etched individually, these contributions could vary from chip to chip and mimic a dose effect. Their constancy shows that they do not change with the O$_2$ exposure. The dose effect resides in the material alone. Ellipsometry before the plasma step confirms that the four films had nearly the same optical and free-carrier response, so the differences after treatment are caused by the treatment alone~\cite{Teke2026}.

\subsection{Low-temperature switching currents}

The normal state thus changes only by scattering added at the surface. The same surface layer could also reduce the cross section that carries the supercurrent. Neither the current densities at 4.2\,K nor the efficiencies $\eta$ show a degradation with the dose [Fig.~\ref{fig:icjc}(a), Table~\ref{tab:eta}]. The larger $\eta$ of D\_60s reflects its reduced ceiling. That of C\_40s reflects its fitted amplitude, the largest of the series. Since the optical resistivity averages over the film including the oxidized surface, as noted above, the calculated ceilings of the treated samples are lower bounds and their efficiencies upper bounds, so the true values lie even closer to that of A\_0s.

The efficiencies (Table~\ref{tab:eta}) agree with the width dependence of bridges from the same PE-ALD process, which decreases from approximately 0.55 at 100\,nm to 0.2 at \SI{10}{\micro\metre}~\cite{KnehrDiss}, or approximately 0.43 interpolated at 500\,nm. Those ratios are evaluated at 4.2\,K. Evaluating our wires there as well changes $\eta$ by at most 3\%, so the comparison does not depend on the choice of temperature. Efficiencies of about one half are typical for NbN bridges~\cite{Charaev2017}, and for straight wires they point to vortex entry at the wire edges rather than localized constrictions as the switching mechanism~\cite{Henrich2012}. The plasma treatment thus creates no new weak spots in the current path at low temperature.

The similarity between the four wires is confined to low temperature. Near $T_\mathrm{c}$ the treated wires fall below A\_0s and their supercurrent vanishes earlier [Fig.~\ref{fig:icjc}(a)]. The collapse saturation of D\_60s lies above that of the untreated sample [Fig.~\ref{fig:icjc}(b)]. Since the depairing current scales as $(1-t)^{3/2}$ near $T_\mathrm{c}$ but is governed by $\Delta_0$ at low temperature, a perturbation that lowers the temperature scale of superconductivity, rather than the zero-temperature condensate, reproduces this pattern of low-temperature convergence and near-$T_\mathrm{c}$ divergence. Geometric constrictions and current crowding are thereby excluded. They suppress the critical current predominantly at low temperature and become ineffective as $\xi(T)$ diverges toward $T_\mathrm{c}$~\cite{Henrich2012,Charaev2017}, the mirror image of the observation, and they would moreover reduce $\eta$ with increasing dose, which is not observed. Homogeneous disorder enhancement is quantitatively insufficient as well. The resistivity increase of 20.7\% predicts a $T_\mathrm{c}$ shift well below the sample-to-sample scatter, as shown above, whereas the suppression observed here amounts to at least 0.5\,K in the supercurrent endpoint.

What remains is a surface mechanism. Both compounds of the oxidized surface identified above, Nb$_2$O$_5$ and the niobium oxynitride, are dielectric~\cite{Nico2016,darlinski1987}. The oxidized surface therefore carries no supercurrent and only reduces the cross section, which the low-temperature currents leave unresolved. The oxidation does not end at a sharp interface, however. Angle-resolved XPS suggests a gradual replacement of nitrogen by oxygen on approaching the surface~\cite{Ermolieff1986,jouve1996}. A partially oxidized but still metallic NbN layer with weakened superconductivity is therefore expected between the dielectric oxide and the intact film. For damaged, partially oxidized edges of NbN strips such a layer has been modeled as a normal or weakly superconducting metal with about twice the resistivity of the intact core~\cite{Charaev2017}. In contrast to the dielectric oxide, this metallic layer can couple to the intact film below. A normal or weakly superconducting surface layer reduces the pair amplitude of the adjacent superconductor over the coherence length $\xi(T)$ by the proximity effect~\cite{deGennes1964,Fominov2001}. At 4.2\,K the order parameter of the core is fully developed and $\xi$ is small against the film thickness, so the reduction is negligible. Close to $T_\mathrm{c}$ the order parameter is weak and $\xi(T)$ spans the film, so the same layer suppresses the entire cross section. The thickness of this layer is expected to vary along the wire, since the oxygen penetration follows the local grain structure and the etched wire edges. The resistive transition percolates through the least degraded segments and keeps the midpoint of the intact film, whereas the supercurrent is limited by the weakest cross section and ends earlier. This reconciles the unchanged resistive $T_\mathrm{c}$ with the reduced supercurrent endpoint.

\section{Conclusion}

We studied the effect of a controlled O$_2$ plasma treatment on PE-ALD NbN films with an as-grown thickness of 11.4\,nm and 500\,nm wide nanowires patterned from them. XPS before and after sputtering locates the plasma-induced oxygen in a surface layer, whereas the interior of the film is chemically indistinguishable from the untreated reference. GIXRD shows that the crystalline $\delta$-NbN below the surface remains intact. The total film thickness saturates between 40 and 60\,s of exposure, indicating a self-limiting growth of the surface oxide. The normal-state resistivity rises by 21\% (dc) and 25\% (optical) with the dose, while the critical temperature, the transition width, and the conduction mechanism remain unchanged. The low-temperature transport is equally robust. The switching currents at 4.2\,K and the depairing efficiencies $\eta = 0.39$--$0.46$ show no systematic dependence on the oxygen dose. However, the current density is decreased with increasing oxygen dose, mainly in the vicinity of $T_\mathrm{c}$, where the supercurrent of the treated wires vanishes at least 0.5\,K lower than that of the untreated wire. This temperature-selective suppression is consistent with a partially oxidized surface layer that acts on the intact film by the proximity effect. At these doses the oxygen therefore narrows the temperature range of the supercurrent from higher temperatures but does not systematically reduce the low-temperature current-carrying capacity. In the future, THz measurements will be performed to gain additional insights into the low-frequency electrodynamics of ALD-grown NbN thin films in dependence on the defects introduced by the oxygen plasma treatment. Also, $B_\mathrm{c2}$ measurements will be performed and analyzed with respect to the effect of the oxygen plasma treatment on the diffusion coefficient of the superconducting carriers. Furthermore, the effect of the oxygen plasma treatment on the performance of SNSPDs will be investigated.

\begin{acknowledgments}
This work was supported in the innovation project ``Infrared single-photon detection via superconducting nanowires tailored by disorder'' by the Carl-Zeiss Stiftung as part of CZS Center QPhoton. The XRD facilities are partially supported by European Fond for Regional Development (EFRE) under the Project No. FKZ 2023 HSB 0025. The ALD equipment is supported by the Free State of Thuringia under the number 2025 VFE 0095 and co-financed by European Union funds within the framework of the European Regional Development Fund (ERDF).
\end{acknowledgments}

\appendix

\section{Diffusion coefficient}\label{app:D}

\begin{figure}[t]
  \centering
  \includegraphics[width=\columnwidth]{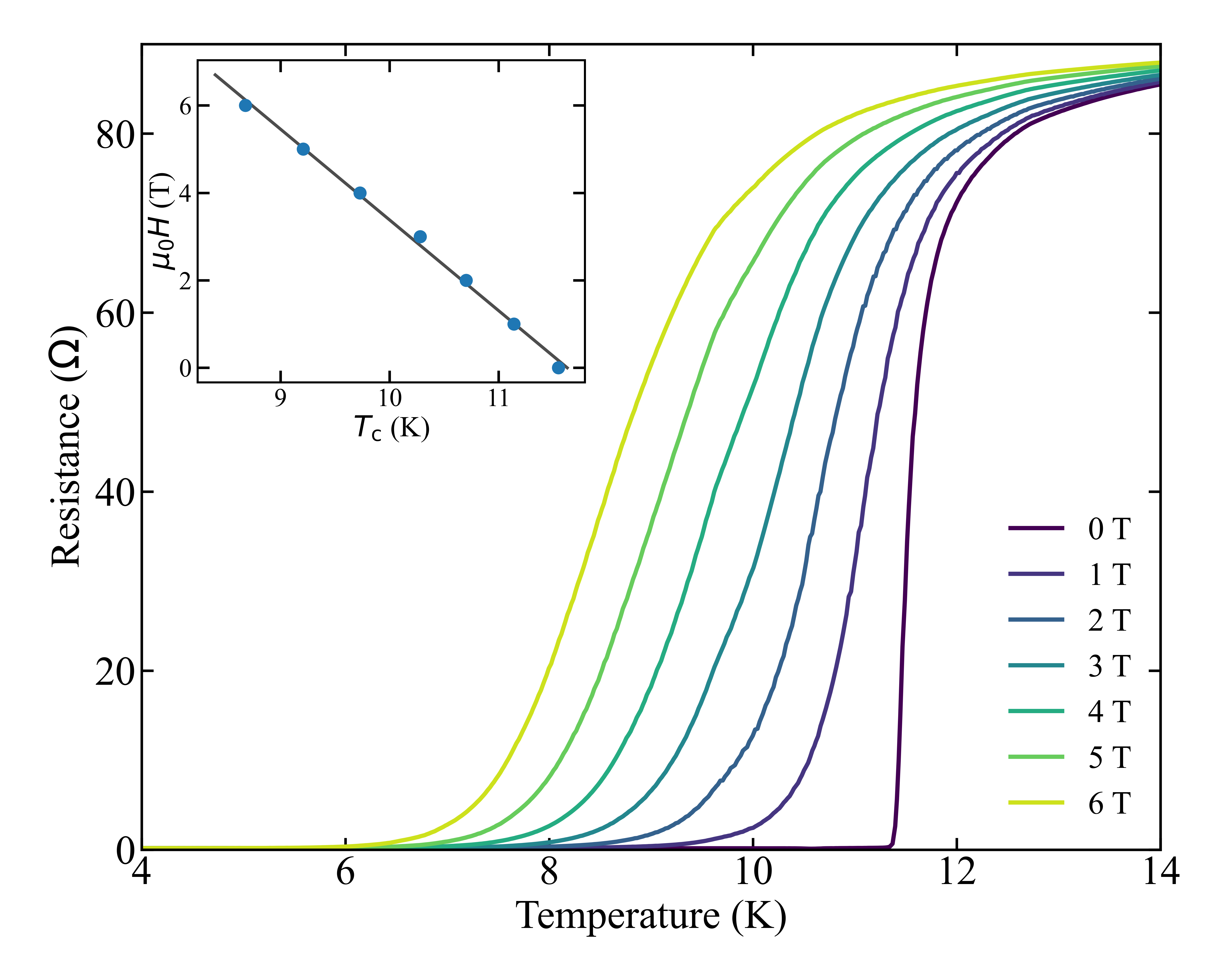}
  \caption{Resistance of the untreated Hall-bar structure as a function of temperature in perpendicular magnetic fields of 0 to 6\,T in steps of 1\,T. Inset: upper critical field from the midpoints of the transitions (symbols) together with a linear fit (line). The slope of $-2.07$\,T/K yields $D = 0.53$\,cm$^2$/s through Eq.~(\ref{eq:whh}).\label{fig:appD}}
\end{figure}

The electron diffusion coefficient of Sec.~\ref{sec:belowTc} was measured on an untreated NbN film deposited with the same recipe, thickness, and substrate stack as sample A\_0s and patterned into a Hall-bar structure with a 130-\si{\micro\metre}-long and 20-\si{\micro\metre}-wide channel. The measurements were performed in a variable-temperature cryostat equipped with a superconducting 6\,T magnet (Cryo Industries of America). The resistance was recorded as a function of temperature at a bias current of \SI{10}{\micro\ampere} in magnetic fields of 0 to 6\,T applied perpendicular to the film plane (Fig.~\ref{fig:appD}). With increasing field the transition shifts to lower temperatures and broadens. The critical temperature at each field was defined as the midpoint of the transition, the same criterion as in Sec.~\ref{sec:methods}. In zero field the midpoint lies at 11.55\,K. Reversing the field at 1\,T changes the midpoint by less than 5\,mK, which confirms that the applied field carries no offset.

The midpoints define the temperature dependence of the upper critical field $B_\mathrm{c2}$ (inset of Fig.~\ref{fig:appD}). It decreases linearly with temperature over the accessible field range. A linear fit over all seven fields yields $\mathrm{d}B_\mathrm{c2}/\mathrm{d}T = -2.07$\,T/K. In the dirty limit the diffusion coefficient follows from this slope alone~\cite{HelfandWerthamer1966,Charaev2017},
\begin{equation}
D = -\frac{4 k_\mathrm{B}}{\pi e}\left(\frac{\mathrm{d}B_\mathrm{c2}}{\mathrm{d}T}\right)^{-1},
\label{eq:whh}
\end{equation}
which gives $D = 0.53$\,cm$^2$/s. The critical temperature itself does not enter Eq.~(\ref{eq:whh}). The 0.6\,K difference between the Hall bar and the wire A\_0s therefore does not affect the result. Reported diffusion coefficients of NbN films with different thicknesses and critical temperatures lie between 0.51 and 0.66\,cm$^2$/s~\cite{Semenov2009}, and a value of 0.56\,cm$^2$/s was obtained for a sputtered film with a critical temperature of 13.55\,K~\cite{Charaev2017}. The film-to-film variation of $D$ is thus small compared with the systematic band considered in Sec.~\ref{sec:discussion}. Extrapolating the fit with $B_\mathrm{c2}(0) = 0.69\,T_\mathrm{c}\,|\mathrm{d}B_\mathrm{c2}/\mathrm{d}T|$~\cite{HelfandWerthamer1966} gives $B_\mathrm{c2}(0) = 16.6$\,T and a coherence length $\xi(0) = \sqrt{\Phi_0/(2\pi B_\mathrm{c2}(0))} = 4.5$\,nm, close to the coherence length of Sec.~\ref{sec:belowTc}.

\bibliography{bib_OygenPlasmaNbN}

\end{document}